\documentclass[
prc,%
aps,%
10pt,%
final,%
notitlepage,%
oneside,%
twocolumn,%
nobibnotes,%
nofootinbib,
superscriptaddress,%
floatfix,%
showkeys,%
showpacs]%
{revtex4-2}

\usepackage{color} 
\usepackage{amsfonts} 
\usepackage{amsbsy} 
\usepackage{mathrsfs}
\usepackage{graphicx}
\usepackage{comment}

\def\lsim{\mathrel{\rlap{ \lower4pt\hbox{\hskip-3pt$\sim$}}
    \raise1pt\hbox{$<$}}} 
\def\gsim{\mathrel{\rlap{ \lower4pt\hbox{\hskip-3pt$\sim$}}
    \raise1pt\hbox{$>$}}} 

\begin{document}

\title{Light Hypernuclei Production in Au+Au Collisions at $\sqrt{s_{NN}}=$ 3 GeV
 within Thermodynamic Approach}

\author{M. Kozhevnikova}\thanks{e-mail: kozhevnikova@jinr.ru}
\affiliation{Veksler and Baldin Laboratory of High Energy Physics,
  JINR Dubna, 141980 Dubna, Russia}
\author{Yu. B. Ivanov}\thanks{e-mail: yivanov@theor.jinr.ru}
\affiliation{Bogoliubov Laboratory of Theoretical Physics, JINR Dubna,
  141980 Dubna, Russia} 
  \affiliation{National Research Center
  "Kurchatov Institute", 123182 Moscow, Russia}

\begin{abstract}

Simulations of the $\Lambda$-hyperon and light-hypernuclei production 
in Au+Au collisions at $\sqrt{s_{NN}}=$ 3 GeV were performed  within updated
Three-fluid Hydrodynamics-based Event Simulator Extended 
by UrQMD (Ultra-relativistic Quantum Molecular Dynamics) final State interactions (THESEUS). 
The light (hyper)nuclei are treated thermodynamically,
i.e. they are considered on the equal basis with hadrons.
The only additional parameter is related to   
the late freeze-out that imitates the afterburner stage for the light (hyper)nuclei
because the UrQMD is not able to dynamically treat them. 
The calculation of hypernuclei production is completely similar to that of light nuclei. 
The hypernuclei results are compared with recent STAR data.  
It is found that the calculated midrapidity $_\Lambda^3$H/$\Lambda$ ratio falls within 
the error bars of the experimental point.  
It is remarkable that large difference between the $t/p$ and $_\Lambda^3$H/$\Lambda$ ratios 
is reproduced without any additional parameters. 
Rapidity distributions of $_\Lambda^3$H/$\Lambda$ and $_\Lambda^4$He/$\Lambda$  ratios are predicted. 
Midrapidity mean transverse momenta of  
protons, $\Lambda$s and light (hyper)nuclei in  central collisions 
well agree with the data. 
The calculated directed flow also reasonably well reproduces of the data.
\end{abstract}
\maketitle

\section{Introduction}

Hypernuclei are an important topic of nuclear physics.
Heavy-ion reactions at relativistic energies
are an abundant source of strangeness and therefore well
suited for the production of light hypernuclei.
The interest in studying hypernuclei in collisions of heavy ions is twofold.
Firstly, the heavy-ion experiments give us information on lifetimes and binding energies of light hypernuclei, 
see, e.g., \cite{STAR:2017gxa,ALICE:2019vlx,STAR:2021orx,STAR:2022zrf}. 
This allow us to refine out understanding of 
hyperon-nucleon interactions and the role of flavor symmetry that are relevant for nuclear structure
\cite{Gal:2016boi,Knoll:2023mqk,Le:2023bfj} 
and astrophysics \cite{Lonardoni:2014bwa,Maslov:2015msa,Fortin:2017cvt}, as well as  
for construction of the hadronic equation of state (EoS) for applications to  
heavy-ion collisions \cite{Khvorostukhin:2006ih}. 
Another aspect of studying production of hypernuclei is directly related to diagnostics of the 
quark-gluon plasma (QGP) formation in heavy-ion collisions. 
It was suggested that the strangeness population factor $S_3 = {_\Lambda^3}H/{^3}He\cdot{p/\Lambda}$
can serve as a probe of the baryon number and strangeness correlation in
the produced matter because of its different behaviors in the QGP and hadronic matter 
\cite{Zhang:2009ba,Steinheimer:2012tb,Shao:2020lbq}. 

In this paper we focus on discussing the mechanism of hypernuclei formation in heavy-ion collisions.
Similarly to the light-nuclei production, alternative mechanisms 
\cite{Steinheimer:2012tb,Aichelin:2019tnk,Glassel:2021rod,Feng:2021rvl,Reichert:2022mek,Buyukcizmeci:2023azb}
are still actively debated. 
The coalescence \cite{Zhang:2009ba,Steinheimer:2012tb,Feng:2021rvl,Reichert:2022mek} and the thermodynamic models 
\cite{Andronic:2010qu,Andronic:2017pug,Steinheimer:2012tb,Reichert:2022mek,Buyukcizmeci:2023azb,Kozhevnikova:2020bdb,Kozhevnikova:2022wms,Kozhevnikova:2023mnw} 
are two of the most popular alternative approaches. 
The model of Parton-Hadron Quantum Molecular Dynamics (PHQMD)  \cite{Aichelin:2019tnk,Glassel:2021rod} 
is based on specific procedures of 
recognition of light (hyper)nuclei. 
Light (hyper)nuclei in this model are not dynamic objects, but rather are associated with 
relatively stable cluster-like correlations. Therefore, 
this approach can be viewed as kinetics of propagation of the correlations. 
The light nuclei act as dynamic objects in kinetic models of Refs. 
\cite{Oliinychenko:2018ugs,Staudenmaier:2021lrg,Sun:2021dlz} (only deuterons) 
\cite{Sun:2022xjr} (all light nuclei up to $^4$He). 
However, these dynamical treatments have not been extended to hypernuclei so far.

As found in Refs. \cite{Steinheimer:2012tb,Reichert:2022mek}, 
both the coalescence and thermodynamic models agree in their predictions 
for the yields of the light (hyper)nuclei.
The thermodynamical approach has an important advantage. 
It does not need additional parameters for the light-(hyper)nuclei treatment.  
It describes the light nuclei on equal basis with hadrons,
i.e. in terms of temperatures and chemical potentials. 
Therefore, its predictive power is the same for light nuclei and hadrons. 
This approach was first realized within the statistical model \cite{Andronic:2010qu}. 
The statistical model gave a good description of  
even hypernuclei and anti-nuclei \cite{Andronic:2017pug}.

Recently data on light-hypernuclei production in Au+Au collisions at $\sqrt{s_{NN}}=$ 3 GeV were 
appeared \cite{STAR:2021orx,Ji:2023fqp,Ji:2023talk,STAR:2022fnj}, some of them 
\cite{Ji:2023fqp,Ji:2023talk} preliminary. 
These data were analyzed in Refs. 
\cite{Buyukcizmeci:2023azb,Reichert:2022mek,Glassel:2021rod}. 
Simulations were  performed within the PHQMD  
approach \cite{Glassel:2021rod}, 
UrQMD+coalescence and UrQMD-hybrid+coalescence approaches were applied in Ref. 
\cite{Reichert:2022mek}, 
JAM+coalescence model (Jet AA microscopic transport Model \cite{Nara:1999dz,Isse:2005nk}) 
was used in STAR papers and presentations \cite{STAR:2021orx,Ji:2023fqp,Ji:2023talk,STAR:2022fnj}.
These data were also analyzed within 
the hybrid dynamical-statistical approach \cite{Buyukcizmeci:2023azb}, which is a modification of 
the Statistical Multifragmentation Model \cite{Bondorf:1995ua}.

In the present paper, we extend the light-nuclei treatment of our previous papers 
\cite{Kozhevnikova:2020bdb,Kozhevnikova:2022wms,Kozhevnikova:2023mnw} 
to hypernuclei production at $\sqrt{s_{NN}}=$ 3 GeV.
This treatment is based on the updated THESEUS generator \cite{Kozhevnikova:2020bdb}, in 
which the light (hyper)nuclei production is considered within the 
thermodynamic treatment.
Our approach is similar to the thermal production from UrQMD hybrid model in Ref. \cite{Steinheimer:2012tb}.
We calculate some bulk properties and directed  flow of protons, $\Lambda$s and light (hyper)nuclei 
and compare them with available STAR data. 
Since the aforementioned coalescence approaches provide a quite reasonable description of the bulk properties, we address the question whether a similar description can be achieved by less demanding means, i.e. using thermodynamics.

\section{The THESEUS Generator} 
  \label{updated THESEUS}

The THESEUS event generator \cite{Batyuk:2016qmb,Batyuk:2017sku}
is based on the  model of the three-fluid dynamics (3FD) \cite{Ivanov:2005yw,Ivanov:2013wha}
complemented by the UrQMD~\cite{Bass:1998ca} for the afterburner stage.
The output of the 3FD model consists of fields of 
local flow velocities and thermodynamic quantities defined on the freeze-out hypersurface.  
The THESEUS generator transforms the 3FD output
into a set of observed particles, i.e. performs the particlization. 
The particlization is followed by the UrQMD afterburning stage.

The initial version of the THESEUS \cite{Batyuk:2016qmb,Batyuk:2017sku} produced 
the primordial nucleons and hyperons, 
i.e. both observable nucleons and those bound in the light (hyper)nuclei. 
These nucleons and hyperons were intended for 
the subsequent use in the coalescence model,   
just as it was done for the light-nuclei production in Refs. 
\cite{Ivanov:2005yw,Ivanov:2017nae}. 
After the coalescence, the nucleons 
and hyperons bound in the light (hyper)nuclei should be   
subtracted from the primordial ones. 
If both nucleons/hyperons and light (hyper)nuclei are sampled from the 3FD output,  
i.e. temperature and chemical potential fields obtained from the EoS, 
in which the light (hyper)nuclei are not included, the bound nucleons/hyperons
become double counted. 
This leads to an overestimation of the total baryon charge in the final state
containing both baryons and clusters. 
To avoid this double counting in the updated version of THESEUS \cite{Kozhevnikova:2020bdb},   
the baryon chemical potentials are recalculated proceeding from the local
baryon number conservation in the system of hadrons extended by
the light-(hyper)nuclei species listed in Tab. \ref{tab:clusters}.
The list of the light (hyper)nuclei includes stable nuclei  
[deuterons ($d$), tritons ($t$), helium isotopes $^3$He and $^4$He], 
and low-lying $^4$He resonances decaying into  stable species  \cite{Shuryak:2019ikv}. 
As compared to our previous papers \cite{Kozhevnikova:2020bdb,Kozhevnikova:2022wms,Kozhevnikova:2023mnw}, 
here we additionally include the stable (with respect to strong decays) 
hypernuclei $_\Lambda^3$H and $_\Lambda^4$He.
The anti-(hyper)nuclei are also included. 
\begin{table}[ht]
\begin{center}
\begin{tabular}{|c|c|c|}
\hline
Nucleus($E$[MeV]) & $J$ &  decay modes, in \% \\
\hline
\hline
$d$           & $1$ & Stable \\
$t$           & $1/2$ & Stable \\
$^3$He        & $1/2$ & Stable \\
$^4$He        & $0$ & Stable \\
$^4$He(20.21) & $0$ & $p$ = 100\\
$^4$He(21.01) & $0$ & $n$ = 24,  $p$ = 76\\
$^4$He(21.84) & $2$ & $n$ = 37,  $p$ = 63  \\
$^4$He(23.33) & $2$ & $n$ = 47,  $p$ = 53  \\
$^4$He(23.64) & $1$ & $n$ = 45,  $p$ = 55 \\
$^4$He(24.25) & $1$ & $n$ = 47,  $p$ = 50,  $d$ = 3 \\
$^4$He(25.28) & $0$ & $n$ = 48,  $p$ = 52\\
$^4$He(25.95) & $1$ & $n$ = 48,  $p$ = 52 \\
$^4$He(27.42) & $2$ & $n$ = 3,   $p$ = 3,   $d$ = 94 \\
$^4$He(28.31) & $1$ & $n$ = 47,  $p$ = 48,  $d$ = 5 \\
$^4$He(28.37) & $1$ & $n$ = 2,   $p$ = 2,   $d$ = 96 \\
$^4$He(28.39) & $2$ & $n$ = 0.2, $p$ = 0.2, $d$ = 99.6 \\
$^4$He(28.64) & $0$ & $d$ = 100 \\
$^4$He(28.67) & $2$ & $d$ = 100 \\
$^4$He(29.89) & $2$ & $n$ = 0.4, $p$ = 0.4, $d$ = 99.2 \\
$_\Lambda^3$H        & $1/2$ & Stable \\
$_\Lambda^4$He       & $0$ & Stable \\
\hline
\end{tabular}
\caption{Stable light (hyper)nuclei and low-lying resonances of the~$^4$He system (from BNL properties of nuclides 
\cite{www.nndc.bnl}).
$J$ denotes the total angular momentum. 
The last column represents branching ratios of the decay channels, in per cents. The 
$p,n,d$ correspond to the emission of protons, neutrons, or deuterons, respectively.
The hypernuclei are replicated from Refs. \cite{Steinheimer:2012tb,Davis:2005mb}. 
}
\label{tab:clusters}
\end{center}
\end{table}

The light (hyper)nuclei are included 
on equal basis with other hadrons in the updated THESEUS 
\cite{Kozhevnikova:2020bdb}. 
They are sampled similarly to other hadrons.
However, there is an important difference 
between treatments of light (hyper)nuclei and other hadrons. 
While the hadrons pass through the UrQMD afterburner stage after
the particlization, the light (hyper)nuclei do not, because the
UrQMD is not able to treat them.
To partially overcome this shortcoming we 
imitate the afterburner for light (hyper)nuclei by late freeze-out in the 3FD,  
following the recipe of Ref. \cite{Kozhevnikova:2020bdb}.

The calculation of hypernuclei production is completely similar to that of light nuclei 
in Ref. \cite{Kozhevnikova:2023mnw}. In Ref. \cite{Kozhevnikova:2023mnw} 
it was found that the late freeze-out, characterized by the freeze-out energy density 
$\varepsilon_{\rm frz}=$ 0.2 GeV/fm$^3$, is preferable for deuterons, tritons, and $^3$He.  
We use precisely the same freeze-out for the calculation of the $_\Lambda^3$H production. 
It was also found \cite{Kozhevnikova:2023mnw} that 
the $^4$He observables are better reproduced with the standard 3FD freeze-out, 
$\varepsilon_{\rm frz}=$ 0.4 GeV/fm$^3$, which indicates  
that the $^4$He nuclei better survive 
in the afterburner stage as more spatially compact and tightly bound objects. 
We use this standard 3FD freeze-out for simulations of the $_\Lambda^4$He production. 
However, the binding energy of $_\Lambda^4$He ($B_\Lambda\simeq$ 2.4 MeV \cite{STAR:2022zrf})
is similar to that of $^3$He ($B_N=$ 2.6 MeV), 
which may imply that the late freeze-out is more suitable for the $_\Lambda^4$He production.
Therefore, the calculations with the late freeze-out are also presented for $_\Lambda^4$He. 
Details of the freeze-out procedure in the 3FD are described in Refs. \cite{Russkikh:2006aa,Ivanov:2008zi}.  
Similarly to Ref. \cite{Kozhevnikova:2023mnw}
three different equations of state (EoS's) are used
in the simulations: 
a purely hadronic EoS \cite{gasEOS} (hadr. EoS) and two EoS's
with deconfinement \cite{Toneev06}, i.e. 
an EoS with a first-order phase transition (1PT EoS) and one with a
smooth crossover transition (crossover EoS).

\section{Bulk Observables} 
  \label{Bulk Observables}

\begin{figure}[!tbh]
  \includegraphics[width=.4\textwidth]{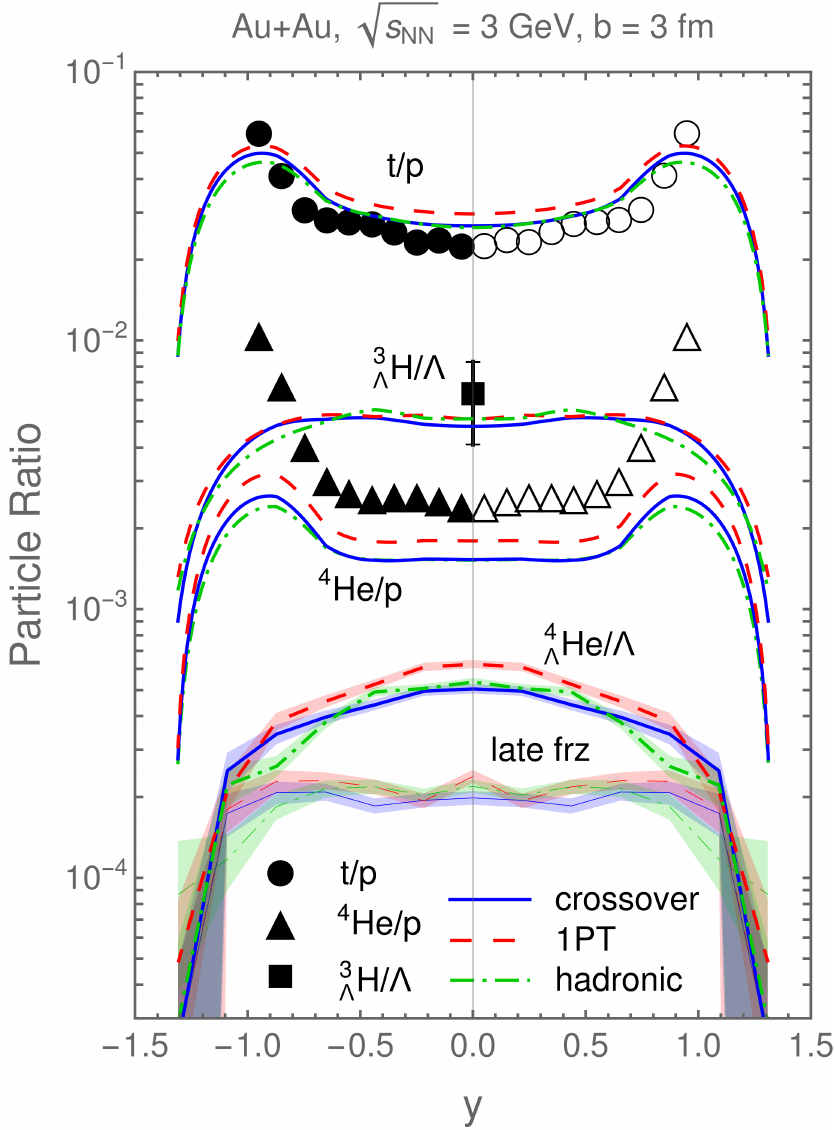}
  \caption{
Rapidity distributions of ratios $t/p$, $^4$He$/p$, $_\Lambda^3$H/$\Lambda$, and
$_\Lambda^4$He/$\Lambda$ in central ($b=$ 3 fm)
Au+Au collisions at collision energy of $\sqrt{s_{NN}}=$ 3 GeV. 
Results are calculated with hadronic, 1PT and crossover EoS's. 
The $t$ and $_\Lambda^3$H yields are calculated with the 
late freeze-out, while $^4$He and
$_\Lambda^4$He ones, with the conventional 3FD freeze-out 
(bold lines). 
Results for $_\Lambda^4$He with the late freeze-out are also displayed (thin lines marked as ``late frz'').
Protons and $\Lambda$s are calculated with the conventional 3FD freeze-out 
and the subsequent UrQMD afterburner.
STAR data for protons and light nuclei (centrality 0--10\%) are from Ref. \cite{STAR:2023uxk}.
The $_\Lambda^3$H/$\Lambda$ midrapidity point is taken from Refs. \cite{Ji:2023talk}.
Full symbols display measured experimental points, whereas the open ones are those reflected with respect
to the midrapidity.}	
    \label{fig:ParticleRatioHyper_y_interp.pdf}
\end{figure}

Rapidity distributions of ratios $t/p$, $^4$He$/p$, $_\Lambda^3$H/$\Lambda$, and
$_\Lambda^4$He/$\Lambda$ in central ($b=$ 3 fm)
Au+Au collisions at collision energy of $\sqrt{s_{NN}}=$ 3 GeV  
are presented in Fig. \ref{fig:ParticleRatioHyper_y_interp.pdf}. 
The proton and $\Lambda$ distributions are calculated within full THESEUS, 
i.e. with the standard 3FD freeze-out and the UrQMD afterburner. 
The $t$ and $_\Lambda^3$H yields are calculated with the 
late freeze-out ($\varepsilon_{\rm frz}=$ 0.2 GeV/fm$^3$), while $^4$He and
$_\Lambda^4$He ones, with the conventional 3FD freeze-out 
($\varepsilon_{\rm frz}=$ 0.4 GeV/fm$^3$) without the subsequent afterburner. 
Since the binding energy of $_\Lambda^4$He is similar to that of $^3$He, 
we also present calculations for $_\Lambda^4$He production with the late freeze-out.
Statistical errors of the $_\Lambda^4$He calculations are displayed by the respective bands.

The strange particles are rare probes at this collision energy. 
The canonical ensemble with exact strangeness conservation is needed for 
their description. The calculated yields of $\Lambda$s,  $_\Lambda^3$H and $_\Lambda^4$He
are considerably overestimated because of grand canonical 
ensemble used in 3FD. Therefore, we do not demonstrate 
the $\Lambda$,  $_\Lambda^3$H and $_\Lambda^4$He yields by themselves. This grand-canonical 
overestimation is canceled in the $_\Lambda^3$H/$\Lambda$ and
$_\Lambda^4$He/$\Lambda$  ratios.

As seen from Fig. \ref{fig:ParticleRatioHyper_y_interp.pdf}, the calculation reasonably well 
reproduces the non-strange ratios. The slight overestimation of the $t/p$ ratio is because of 
underestimation of the proton yield, see Ref. \cite{Kozhevnikova:2023mnw}. The $^4$He 
is underestimated in our calculations. Apparently, inclusion of light-nuclei resonances with $A=5$ into the scheme, i.e. those of $^5$H, $^5$He, and $^5$Li \cite{Vovchenko:2020dmv} that  
decay into $^4$He, may correct this underestimation. 
The calculated result for the midrapidity $_\Lambda^3$H/$\Lambda$ ratio falls within 
the error bars of the experimental point \cite{Ji:2023talk}.  
It is remarkable that large difference between the $t/p$ and $_\Lambda^3$H/$\Lambda$ ratios 
is reproduced without any additional parameters. 

The collision process develops in the hadronic phase in all considered scenarios, 
i.e. hadronic, 1PT and crossover ones. The corresponding EoS's are very similar in the 
hadronic phase but not identical. Therefore, EoS-induced differences in the ratios indicate 
uncertainties of the model predictions.

\begin{figure}[!tbh]
  \includegraphics[width=.39\textwidth]{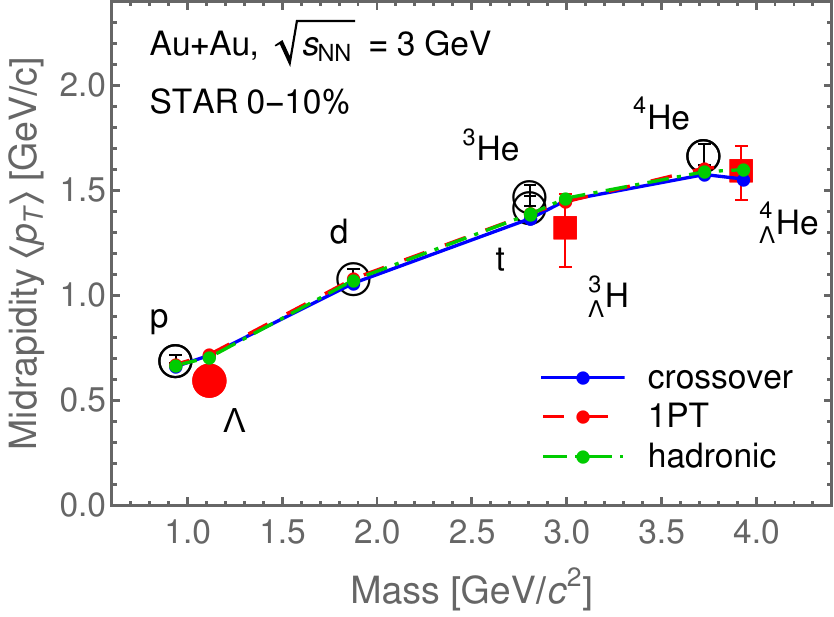}
  \caption{
Midrapidity mean transverse momentum of  
protons, $\Lambda$ hyperons and light (hyper)nuclei in  central ($b=$ 3 fm)
Au+Au collisions at collision energy of $\sqrt{s_{NN}}=$ 3 GeV.
Results are calculated with hadronic, 1PT and crossover EoS's. 
Protons are calculated within the conventional 3FD freeze-out with the subsequent UrQMD afterburner.
Deuterons, tritons, $^3$He, and $_\Lambda^3$H/$\Lambda$ are calculated with the 
late freeze-out, 
while $^4$He and $_\Lambda^4$He, the   
conventional 3FD freeze-out.  
STAR data are from Refs. \cite{Ji:2023fqp,STAR:2023uxk}.}
    \label{mid_pT_average.pdf}
\end{figure}

Midrapidity mean transverse momentum of  
protons, $\Lambda$s and light (hyper)nuclei in  central ($b=$ 3 fm) collisions is displayed in 
Fig. \ref{mid_pT_average.pdf}. This quantity characterizes the radial flow. 
Curves in Fig. \ref{mid_pT_average.pdf} are displayed only for eye guidance.
As seen, these curves (for three different EoS's) in fact coincide. 
Results of the calculations are shown by dots on these curves. 
Moreover, results for strange and non-strange species lie on the same curves. 
The calculated points well agree with the data \cite{Ji:2023fqp,STAR:2023uxk}. 
Even slight deviation of these curves from straight lines is reproduced.

\section{Directed flow} 
  \label{Directed flow}

\begin{figure}[!tbh]
\includegraphics[width=.47\textwidth]{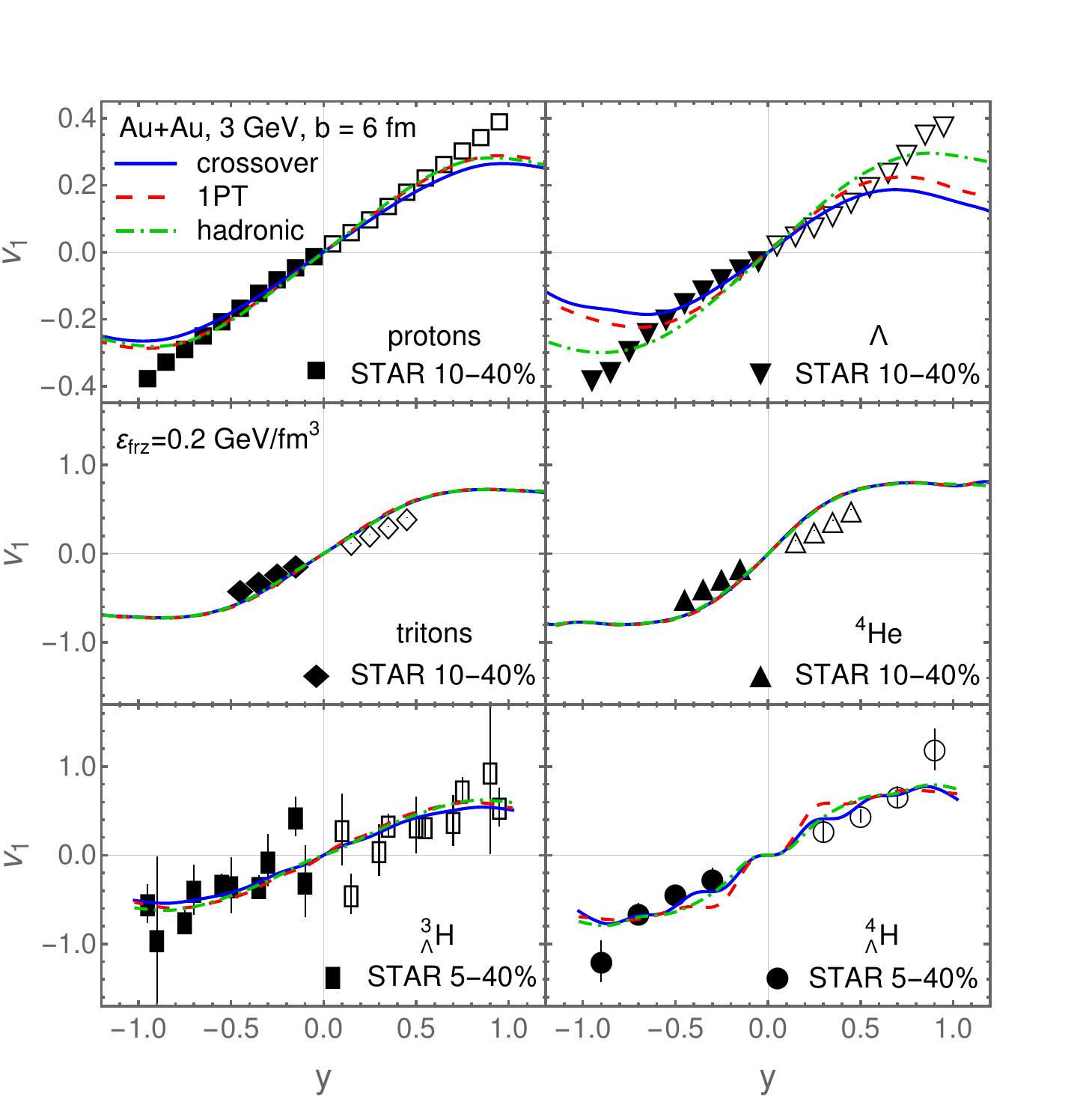}
  \caption{
Directed flow of  
protons, $\Lambda$ hyperons and light (hyper)nuclei 
(tritons, $^4$He, $_\Lambda^3$H, and $_\Lambda^4$He) 
as function of rapidity in semicentral ($b=$ 6 fm)    
Au+Au collisions at collision energy of $\sqrt{s_{NN}}=$ 3 GeV. 
Results are calculated 
with hadronic, 1PT and crossover EoS's. 
The THESEUS simulations for light (hyper)nuclei were performed with the 
late freeze-out ($\varepsilon_{\rm frz}=$ 0.2 GeV/fm$^3$). 
The flow of protons and $\Lambda$s is calculated with the conventional 3FD freeze-out 
followed by the subsequent UrQMD afterburner.
STAR data are from Refs. \cite{STAR:2022fnj,STAR:2021ozh,STAR:2021yiu}.
Full symbols display measured experimental points, whereas the open ones are those reflected with respect
to the midrapidity.}	
    \label{v1_hyper_frz02.pdf}
\end{figure}

The directed flow is a more delicate observable. 
The calculated directed flow of protons, $\Lambda$ hyperons and light (hyper)nuclei
(tritons, $^4$He, $_\Lambda^3$H and $_\Lambda^4$He) 
as function of rapidity in semicentral ($b=$ 6 fm)    
Au+Au collisions at collision energy of $\sqrt{s_{NN}}=$ 3 GeV is presented in 
Fig. \ref{v1_hyper_frz02.pdf}. 
The results are compared with STAR data \cite{STAR:2022fnj,STAR:2021ozh,STAR:2021yiu}. 
We do not display results for light nuclei (deuterons and $^3$He)
because they are not directly related to the considered hypernuclei. 
Results for all light nuclei can be found in Ref. \cite{Kozhevnikova:2023mnw}. 
The THESEUS simulations for light (hyper)nuclei are performed for the 
late freeze-out ($\varepsilon_{\rm frz}=$ 0.2 GeV/fm$^3$) for three EoS's. 
The flow of protons and $\Lambda$ hyperons is calculated within the full THESEUS, 
i.e. with the conventional 3FD freeze-out and the subsequent UrQMD afterburner.

\begin{figure}[!tbh]
\includegraphics[width=.47\textwidth]{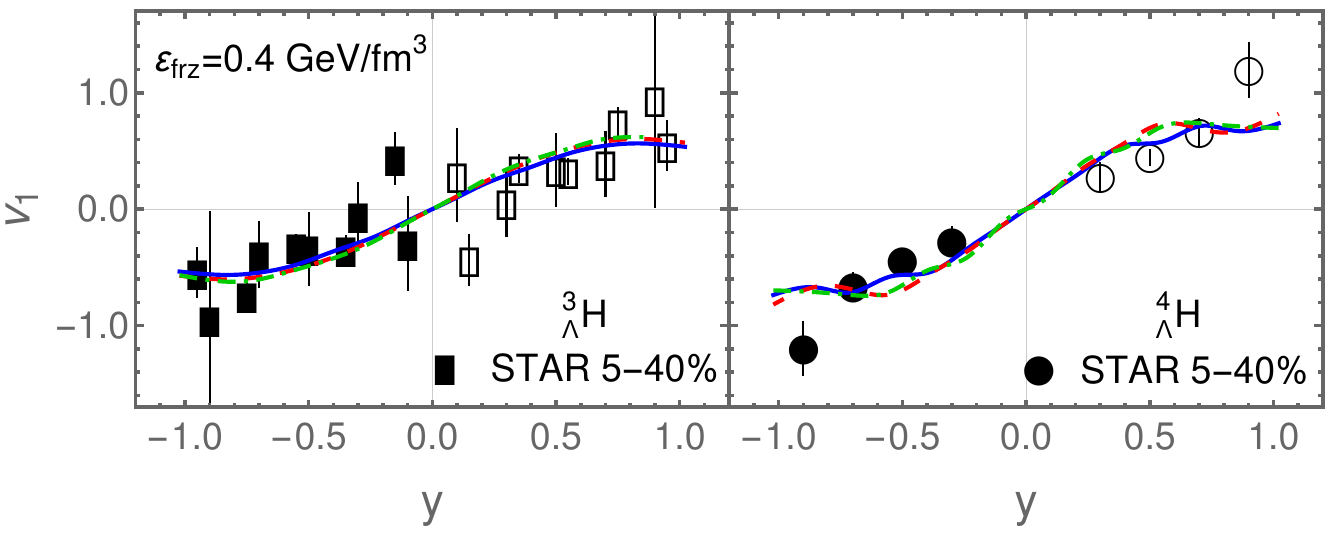}
  \caption{
The same as in Fig. \ref{v1_hyper_frz02.pdf} but only for light hypernuclei
($_\Lambda^3$H and $_\Lambda^4$He) calculated with the conventional 3FD freeze-out
($\varepsilon_{\rm frz}=$ 0.4 GeV/fm$^3$).}  
    \label{v1_hyper_frz04.pdf}
\end{figure}

The directed proton flow is almost independent of the used EoS \cite{Kozhevnikova:2023mnw}. 
The calculated results perfectly 
(except for very forward and backward rapidities) reproduce the experimental 
proton flow \cite{STAR:2021yiu}. Agreement with the data \cite{STAR:2021ozh}
becomes worse with increase of atomic number of light nucleus. 
If the calculated midrapidity slope of the  triton directed flow is only slightly steeper than 
the experimental one, for $^4$He it is already noticeably steeper.

The directed $\Lambda$ flow depends on the EoS, while that of hypernuclei is again EoS independent
(up to the statistical fluctuations). 
Apparently the nucleon content of the hypernuclei dominates in the $v_1$ formation. 
The crossover scenario results in the best in reproduction of the midrapidity slope of the $\Lambda$ flow. 
The $_\Lambda^4$He flow is reproduced to the same extent as that of light nuclei. 
It is difficult to judge the degree of agreement with the $_\Lambda^3$H flow data because of 
their large error bars.

In Fig. \ref{v1_hyper_frz02.pdf}, the directed flow of $^4$He and $_\Lambda^4$He is calculated with 
the late freeze-out ($\varepsilon_{\rm frz}=$ 0.2 GeV/fm$^3$) instead of the conventional 3FD freeze-out
that is preferable for $^4$He and presumably for $_\Lambda^4$He. 
The reason is that the $^4$He directed flow is independent of the type (late or conventional) of 
the used freeze-out, as demonstrated in Ref. \cite{Kozhevnikova:2023mnw}. 
Nevertheless, we additionally checked this independence for the $_\Lambda^4$He. 
The  results of calculation of $v_1$ of  $_\Lambda^3$H and $_\Lambda^4$He with 
the conventional 3FD freeze-out is presented in Fig. \ref{v1_hyper_frz04.pdf}. 
As seen, the $v_1$ flows with the conventional freeze-out for both hypernuclei are just identical 
(up to statistical fluctuations) to the late-freeze-out ones. 
The midrapidity slope of the proton flow also remains unchanged after the afterburner \cite{Kozhevnikova:2023mnw}.
All this indicates that the baryon directed flow is formed at the early stage of the reaction.

\section{Summary}
\label{Summary}

Simulations of the $\Lambda$-hyperon and light-hypernuclei production 
in Au+Au collisions at $\sqrt{s_{NN}}=$ 3 GeV were performed
within the updated THESEUS event generator \cite{Kozhevnikova:2020bdb}. 
In the updated THESEUS, the light (hyper)nuclei are treated thermodynamically,
i.e. they are considered on the equal basis with hadrons.
The only additional parameter associated with the light (hyper)nuclei is related to   
the late freeze-out that imitates the afterburner stage because 
the UrQMD is not able to dynamically treat the light (hyper)nuclei. 
This is a less demanding way to describe the light-(hyper)nuclei production as compared to 
the coalescence.

The calculation of hypernuclei production is completely similar to that of light nuclei 
in Ref. \cite{Kozhevnikova:2023mnw}. 
In Ref. \cite{Kozhevnikova:2023mnw} 
it was found that the late freeze-out
is preferable for deuterons, tritons, and $^3$He.  
We used precisely the same freeze-out for the calculation of the $_\Lambda^3$H production. 
It was also found \cite{Kozhevnikova:2023mnw} that 
the $^4$He observables are better reproduced with the standard 3FD freeze-out,
which indicates that the $^4$He nuclei better survive 
in the afterburner stage as more spatially compact and tightly bound objects. 
We used this standard 3FD freeze-out for simulations of the $_\Lambda^4$He production. 
However, the binding energy of $_\Lambda^4$He ($B_\Lambda\simeq$ 2.4 MeV \cite{STAR:2022zrf})
is similar to that of $^3$He ($B_N=$ 2.6 MeV), 
which may imply that the late freeze-out is more suitable for the $_\Lambda^4$He production.
The hypernuclei results were compared with recent STAR data 
\cite{Ji:2023fqp,Ji:2023talk,STAR:2022fnj}, as well as with the 
results of light-nuclei calculations \cite{Kozhevnikova:2023mnw}.

It is found that the calculated midrapidity $_\Lambda^3$H/$\Lambda$ ratio falls within 
the error bars of the experimental point \cite{Ji:2023talk}.  
It is remarkable that large difference between the $t/p$ and $_\Lambda^3$H/$\Lambda$ ratios 
is reproduced without any additional parameters. 
Rapidity distributions of $_\Lambda^3$H/$\Lambda$ and $_\Lambda^4$He/$\Lambda$  ratios are predicted. 
Midrapidity mean transverse momenta of  
protons, $\Lambda$s and light (hyper)nuclei in  central collisions 
well agree with the data \cite{Ji:2023fqp,Ji:2023talk}. 
The calculated directed flow also reasonably well reproduces of the data \cite{STAR:2022fnj}.
The directed flow turned out to be independent of the type (late or
conventional) of the freeze-out. 
This indicates that the baryon directed flow is formed at the early stage of the reaction.

\begin{acknowledgments}

We are sincerely grateful to Iurii Karpenko and David Blaschke 
who made enormous contributions  on the early stage of this project.  
Fruitful discussions with  D.N. Voskresensky are gratefully acknowledged.
This work was carried out using computing resources of the federal collective usage center 
``Complex for simulation and data processing for mega-science facilities'' 
at NRC "Kurchatov Institute" \cite{ckp.nrcki.ru} 
and computing resources of the supercomputer "Govorun" at JINR \cite{govorun}. 

\end{acknowledgments}



\begin{thebibliography}{999}
%
%
%
\bibitem{STAR:2017gxa}
L.~Adamczyk \textit{et al.} [STAR],
Measurement of the $^3_{\Lambda}$H lifetime in Au+Au collisions at the BNL Relativistic Heavy Ion Collider,
Phys. Rev. C \textbf{97}, no.5, 054909 (2018)
doi:10.1103/PhysRevC.97.054909
[arXiv:1710.00436 [nucl-ex]].
%
\bibitem{ALICE:2019vlx}
S.~Acharya \textit{et al.} [ALICE],
$^3_\Lambda\mathrm{H}$ and $^3_{\bar{\Lambda}}\mathrm{\overline{H}}$ lifetime measurement in Pb-Pb collisions at $\sqrt{s_{\mathrm{NN}}} = $ 5.02 TeV via two-body decay,
Phys. Lett. B \textbf{797}, 134905 (2019)
[arXiv:1907.06906 [nucl-ex]].
%
\bibitem{STAR:2021orx}
M.~Abdallah \textit{et al.} [STAR],
Measurements of $H_\Lambda^3$ and $H_\Lambda^4$ Lifetimes and Yields in Au+Au Collisions in the High Baryon Density Region,
Phys. Rev. Lett. \textbf{128}, no.20, 202301 (2022)
doi:10.1103/PhysRevLett.128.202301
[arXiv:2110.09513 [nucl-ex]].
%
\bibitem{STAR:2022zrf}
M.~Abdallah \textit{et al.} [STAR],
Measurement of H\ensuremath{\Lambda}4 and He\ensuremath{\Lambda}4 binding energy in Au+Au collisions at sNN = 3 GeV,
Phys. Lett. B \textbf{834}, 137449 (2022)
doi:10.1016/j.physletb.2022.137449
[arXiv:2207.00778 [nucl-ex]].
%
%
%
\bibitem{Gal:2016boi}
A.~Gal, E.~V.~Hungerford and D.~J.~Millener,
Strangeness in nuclear physics,
Rev. Mod. Phys. \textbf{88}, no.3, 035004 (2016)
[arXiv:1605.00557 [nucl-th]].
%
\bibitem{Knoll:2023mqk}
M.~Kn\"oll and R.~Roth,
Hyperon-nucleon interaction constrained by light hypernuclei,
Phys. Lett. B \textbf{846}, 138258 (2023)
[arXiv:2307.11577 [nucl-th]].
%
\bibitem{Le:2023bfj}
H.~Le, J.~Haidenbauer, U.~G.~Mei\ss{}ner and A.~Nogga,
Separation energies of light $\Lambda$ hypernuclei and their theoretical uncertainties,
Eur. Phys. J. A \textbf{60}, no.1, 3 (2024)
doi:10.1140/epja/s10050-023-01219-w
[arXiv:2308.01756 [nucl-th]].
%
%
\bibitem{Lonardoni:2014bwa}
D.~Lonardoni, A.~Lovato, S.~Gandolfi and F.~Pederiva,
Hyperon Puzzle: Hints from Quantum Monte Carlo Calculations,
Phys. Rev. Lett. \textbf{114}, no.9, 092301 (2015)
[arXiv:1407.4448 [nucl-th]].
%
\bibitem{Maslov:2015msa}
K.~A.~Maslov, E.~E.~Kolomeitsev and D.~N.~Voskresensky,
Solution of the Hyperon Puzzle within a Relativistic Mean-Field Model,
Phys. Lett. B \textbf{748}, 369-375 (2015)
[arXiv:1504.02915 [astro-ph.HE]];
%
Making a soft relativistic mean-field equation of state stiffer at high density,
Phys. Rev. C \textbf{92}, no.5, 052801 (2015)
[arXiv:1508.03771 [astro-ph.HE]];
%
Relativistic Mean-Field Models with Scaled Hadron Masses and Couplings: Hyperons and Maximum Neutron Star Mass,
Nucl. Phys. A \textbf{950}, 64-109 (2016)
[arXiv:1509.02538 [astro-ph.HE]].
%
\bibitem{Fortin:2017cvt}
M.~Fortin, S.~S.~Avancini, C.~Provid\^encia and I.~Vida\~na,
Hypernuclei and massive neutron stars,
Phys. Rev. C \textbf{95}, no.6, 065803 (2017)
[arXiv:1701.06373 [nucl-th]].
%
\bibitem{Khvorostukhin:2006ih}
A.~S.~Khvorostukhin, V.~D.~Toneev and D.~N.~Voskresensky,
Equation of State for Hot and Dense Matter: sigma- omega- rho Model with Scaled Hadron Masses and Couplings,
Nucl. Phys. A \textbf{791}, 180-221 (2007)
[arXiv:nucl-th/0612058 [nucl-th]];
%
Relativistic Mean-Field Model with Scaled Hadron Masses and Couplings,
Nucl. Phys. A \textbf{813}, 313-346 (2008)
[arXiv:0802.3999 [nucl-th]].
%
%
\bibitem{Zhang:2009ba}
S.~Zhang, J.~H.~Chen, H.~Crawford, D.~Keane, Y.~G.~Ma and Z.~B.~Xu,
Searching for onset of deconfinement via hypernuclei and baryon-strangeness correlations,
Phys. Lett. B \textbf{684}, 224-227 (2010)
[arXiv:0908.3357 [nucl-ex]].
%
\bibitem{Steinheimer:2012tb}
J.~Steinheimer, K.~Gudima, A.~Botvina, I.~Mishustin, M.~Bleicher and H.~Stocker,
Hypernuclei, dibaryon and antinuclei production in high energy heavy ion collisions: Thermal production versus Coalescence,
Phys. Lett. B \textbf{714}, 85-91 (2012)
[arXiv:1203.2547 [nucl-th]].
%
\bibitem{Shao:2020lbq}
T.~Shao, J.~Chen, C.~M.~Ko, K.~J.~Sun and Z.~Xu,
Yield ratio of hypertriton to light nuclei in heavy-ion collisions from $\rm \sqrt{s_{NN}}$ = 4.9 GeV to 2.76 TeV,
Chin. Phys. C \textbf{44}, no.11, 114001 (2020)
[arXiv:2004.02385 [nucl-ex]].
%
%
%
\bibitem{Aichelin:2019tnk}
J.~Aichelin, E.~Bratkovskaya, A.~Le F\`evre, V.~Kireyeu, V.~Kolesnikov, Y.~Leifels, V.~Voronyuk and G.~Coci,
Parton-hadron-quantum-molecular dynamics: A novel microscopic $n$-body transport approach for heavy-ion collisions, dynamical cluster formation, and hypernuclei production,
Phys. Rev. C \textbf{101}, no.4, 044905 (2020)
[arXiv:1907.03860 [nucl-th]].
%
\bibitem{Glassel:2021rod}
S.~Gl\"a\ss{}el, V.~Kireyeu, V.~Voronyuk, J.~Aichelin, C.~Blume, E.~Bratkovskaya, G.~Coci, V.~Kolesnikov and M.~Winn,
Cluster and hypercluster production in relativistic heavy-ion collisions within the parton-hadron-quantum-molecular-dynamics approach,
Phys. Rev. C \textbf{105}, no.1, 014908 (2022)
[arXiv:2106.14839 [nucl-th]].
%
\bibitem{Feng:2021rvl}
Z.~Q.~Feng,
Dynamics of light hypernuclei in collisions of $^{197}$Au+$^{197}$Au at GeV energies,
Eur. Phys. J. A \textbf{57}, no.1, 18 (2021)
[arXiv:2109.01270 [nucl-th]].
%
\bibitem{Reichert:2022mek}
T.~Reichert, J.~Steinheimer, V.~Vovchenko, B.~D\"onigus and M.~Bleicher,
Energy dependence of light hypernuclei production in heavy-ion collisions from a coalescence and statistical-thermal model perspective,
Phys. Rev. C \textbf{107}, no.1, 014912 (2023)
[arXiv:2210.11876 [nucl-th]].
%
\bibitem{Buyukcizmeci:2023azb}
N.~Buyukcizmeci, T.~Reichert, A.~S.~Botvina and M.~Bleicher,
Nucleosynthesis of light nuclei and hypernuclei in central Au+Au collisions at sNN=3 GeV,
Phys. Rev. C \textbf{108}, no.5, 054904 (2023)
[arXiv:2306.17145 [nucl-th]].
%
%
%
\bibitem{Andronic:2010qu}
A.~Andronic, P.~Braun-Munzinger, J.~Stachel and H.~Stocker,
Production of light nuclei, hypernuclei and their antiparticles in relativistic nuclear collisions,
Phys. Lett. B \textbf{697}, 203-207 (2011)
[arXiv:1010.2995 [nucl-th]].
%
\bibitem{Andronic:2017pug}
A.~Andronic, P.~Braun-Munzinger, K.~Redlich and J.~Stachel,
Decoding the phase structure of QCD via particle production at high energy,
Nature \textbf{561}, no.7723, 321-330 (2018)
[arXiv:1710.09425 [nucl-th]].
%
%
%
\bibitem{Kozhevnikova:2020bdb}
M.~Kozhevnikova, Y.~B.~Ivanov, I.~Karpenko, D.~Blaschke and O.~Rogachevsky,
Update of the Three-fluid Hydrodynamics-based Event Simulator: light-nuclei production in heavy-ion collisions,
Phys. Rev. C \textbf{103}, no.4, 044905 (2021)
[arXiv:2012.11438 [nucl-th]].
%
\bibitem{Kozhevnikova:2022wms}
M.~Kozhevnikova and Y.~B.~Ivanov,
Light-nuclei production in heavy-ion collisions within a thermodynamical approach,
Phys. Rev. C \textbf{107}, no.2, 024903 (2023)
[arXiv:2210.07334 [nucl-th]];
Light-Nuclei Production in Heavy-Ion Collisions at = 6.4 \textendash{} 19.6 GeV in THESEUS Generator Based on Three-Fluid Dynamics,
Particles \textbf{6}, no.1, 440-450 (2023). 
%
\bibitem{Kozhevnikova:2023mnw}
M.~Kozhevnikova and Y.~B.~Ivanov,
Light-nuclei production in Au+Au collisions at sNN=3 GeV within a thermodynamical approach: Bulk properties and collective flow,
Phys. Rev. C \textbf{109}, no.1, 014913 (2024)
[arXiv:2311.08092 [nucl-th]].
%
%
\bibitem{Oliinychenko:2018ugs}
D.~Oliinychenko, L.~G.~Pang, H.~Elfner and V.~Koch,
Microscopic study of deuteron production in PbPb collisions at $\sqrt{s} = 2.76 TeV$ via hydrodynamics and a hadronic afterburner,
Phys. Rev. C \textbf{99}, no.4, 044907 (2019)
[arXiv:1809.03071 [hep-ph]].
%
\bibitem{Staudenmaier:2021lrg}
J.~Staudenmaier, D.~Oliinychenko, J.~M.~Torres-Rincon and H.~Elfner,
Deuteron production in relativistic heavy ion collisions via stochastic multiparticle reactions,
Phys. Rev. C \textbf{104}, no.3, 034908 (2021)
[arXiv:2106.14287 [hep-ph]].
%
\bibitem{Sun:2021dlz}
K.~J.~Sun, R.~Wang, C.~M.~Ko, Y.~G.~Ma and C.~Shen,
Relativistic kinetic approach to light nuclei production in high-energy nuclear collisions,
[arXiv:2106.12742 [nucl-th]].
%
\bibitem{Sun:2022xjr}
K.~J.~Sun, R.~Wang, C.~M.~Ko, Y.~G.~Ma and C.~Shen,
Unveiling the dynamics of little-bang nucleosynthesis,
Nature Commun. \textbf{15}, no.1, 1074 (2024)
[arXiv:2207.12532 [nucl-th]].
%
%
\bibitem{Ji:2023fqp}
Y.~Ji [STAR],
Measurements on the production and properties of light hypernuclei at STAR,
EPJ Web Conf. \textbf{276}, 04003 (2023)
%
\bibitem{Ji:2023talk}
Yuanjing Ji,  talk at Quark Matter 2023, 
\url{https://indico.cern.ch/event/1139644/contributions/5456392/attachments/2707583/4708403/talk_FXT_H3L_Sep08_v11.pdf}
%
\bibitem{STAR:2022fnj}
B.~Aboona \textit{et al.} [STAR],
Observation of Directed Flow of Hypernuclei H$\ensuremath{\Lambda}3 and H\ensuremath{\Lambda}4$ in $\sqrt{s_{NN}}=$ 3 GeV Au+Au Collisions at RHIC,
Phys. Rev. Lett. \textbf{130}, no.21, 212301 (2023)
[arXiv:2211.16981 [nucl-ex]].
%
%
\bibitem{Nara:1999dz}
Y.~Nara, N.~Otuka, A.~Ohnishi, K.~Niita and S.~Chiba,
Study of relativistic nuclear collisions at AGS energies from p + Be to Au + Au with hadronic cascade model,
Phys. Rev. C \textbf{61}, 024901 (2000)
[arXiv:nucl-th/9904059 [nucl-th]].
%
\bibitem{Isse:2005nk}
M.~Isse, A.~Ohnishi, N.~Otuka, P.~K.~Sahu and Y.~Nara,
Mean-field effects on collective flows in high-energy heavy-ion collisions from AGS to SPS energies,
Phys. Rev. C \textbf{72}, 064908 (2005)
[arXiv:nucl-th/0502058 [nucl-th]].
%
\bibitem{Bondorf:1995ua}
J.~P.~Bondorf, A.~S.~Botvina, A.~S.~Ilinov, I.~N.~Mishustin and K.~Sneppen,
Statistical multifragmentation of nuclei,
Phys. Rept. \textbf{257}, 133-221 (1995)
%
%
  \bibitem{Batyuk:2016qmb} P.~Batyuk {\it et al.},
   Event simulation based on three-fluid hydrodynamics for
   collisions at energies available at the Dubna Nuclotron-based Ion
   Collider Facility and at the Facility for Antiproton and Ion
   Research in Darmstadt,
  Phys.\ Rev.\ C {\bf 94}, 044917 (2016)
	[arXiv:1608.00965 [nucl-th]].
%
\bibitem{Batyuk:2017sku}
P.~Batyuk, D.~Blaschke, M.~Bleicher, Y.~B.~Ivanov, I.~Karpenko, L.~Malinina, S.~Merts, M.~Nahrgang, H.~Petersen and O.~Rogachevsky,
Three-fluid Hydrodynamics-based Event Simulator Extended by UrQMD final State interactions (THESEUS) for FAIR-NICA-SPSBES/RHIC energies,
EPJ Web Conf. \textbf{182}, 02056 (2018)
[arXiv:1711.07959 [nucl-th]].
%
%
\bibitem{Ivanov:2013wha}
Y.~B.~Ivanov,
Alternative Scenarios of Relativistic Heavy-Ion Collisions: I. Baryon Stopping,
Phys. Rev. C \textbf{87}, no.6, 064904 (2013)
[arXiv:1302.5766 [nucl-th]].
  %
  \bibitem{Ivanov:2005yw}
  Y.~B.~Ivanov, V.~N.~Russkikh and V.~D.~Toneev,
  Relativistic heavy-ion collisions within 3-fluid hydrodynamics:
   Hadronic scenario,
  Phys.\ Rev.\ C {\bf 73}, 044904 (2006)
	[nucl-th/0503088].
%
\bibitem{Bass:1998ca}
S.~A.~Bass, M.~Belkacem, M.~Bleicher, M.~Brandstetter, L.~Bravina, C.~Ernst, L.~Gerland, M.~Hofmann, S.~Hofmann and J.~Konopka, \textit{et al.},
Microscopic models for ultrarelativistic heavy ion collisions,
Prog. Part. Nucl. Phys. \textbf{41}, 255-369 (1998)
[arXiv:nucl-th/9803035 [nucl-th]].
%
\bibitem{Ivanov:2017nae} 
  Y.~B.~Ivanov and A.~A.~Soldatov,
  Light fragment production at CERN Super Proton Synchrotron,
  Eur.\ Phys.\ J.\ A {\bf 53}, no. 11, 218 (2017)
[arXiv:1703.05040 [nucl-th]].
%
\bibitem{Shuryak:2019ikv}
E.~Shuryak and J.~M.~Torres-Rincon,
Baryon preclustering at the freeze-out of heavy-ion collisions and light-nuclei production,
Phys. Rev. C \textbf{101}, no.3, 034914 (2020)
[arXiv:1910.08119 [nucl-th]].
%
\bibitem{www.nndc.bnl}
\url{https://www.nndc.bnl.gov/nudat3/getdataset.jsp?nucleus=4HE&unc=nds}
%
\bibitem{Davis:2005mb}
D.~H.~Davis,
50 years of hypernuclear physics. I. The early experiments,
Nucl. Phys. A \textbf{754}, 3-13 (2005)
%
%
\bibitem{Russkikh:2006aa} 
  V.~N.~Russkikh and Yu.~B.~Ivanov,
  Dynamical freeze-out in 3-fluid hydrodynamics,
  Phys.\ Rev.\ C {\bf 76}, 054907 (2007)  
[nucl-th/0611094].
%
\bibitem{Ivanov:2008zi} 
  Yu.~B.~Ivanov and V.~N.~Russkikh,
  On freeze-out problem in relativistic hydrodynamics,
  Phys.\ Atom.\ Nucl.\  {\bf 72}, 1238 (2009)  
	[arXiv:0810.2262 [nucl-th]].
%
  %
  \bibitem{gasEOS}
I.~N.~Mishustin, V.~N.~Russkikh and L.~M.~Satarov,
Fluid dynamical model of relativistic heavy ion collision,
Sov. J. Nucl. Phys. \textbf{54}, 260-314 (1991).
%
  \bibitem{Toneev06}
  A.~S.~Khvorostukin, V.~V.~Skokov, V.~D.~Toneev and K.~Redlich,
  Lattice QCD constraints on the nuclear equation of state,
  Eur.\ Phys.\ J.\ C {\bf 48}, 531 (2006)
  [nucl-th/0605069].
%
%
%
\bibitem{STAR:2023uxk}
 [STAR],
Production of Protons and Light Nuclei in Au+Au Collisions at $\sqrt{s_{\mathrm{NN}}}$ = 3 GeV with the STAR Detector,
[arXiv:2311.11020 [nucl-ex]].
%
\bibitem{Vovchenko:2020dmv}
V.~Vovchenko, B.~D\"onigus, B.~Kardan, M.~Lorenz and H.~Stoecker,
Feeddown contributions from unstable nuclei in relativistic heavy-ion collisions,
Phys. Lett. \textbf{B}, 135746 (2020)
[arXiv:2004.04411 [nucl-th]].
%
\bibitem{STAR:2021ozh}
M.~S.~Abdallah \textit{et al.} [STAR],
Light nuclei collectivity from~$\sqrt{s_{NN}}$ = 3 GeV Au+Au collisions at RHIC,
Phys. Lett. B \textbf{827}, 136941 (2022)
[arXiv:2112.04066 [nucl-ex]].
%
\bibitem{STAR:2021yiu}
M.~S.~Abdallah \textit{et al.} [STAR],
Disappearance of partonic collectivity in sNN=3GeV Au+Au collisions at RHIC,
Phys. Lett. B \textbf{827}, 137003 (2022)
[arXiv:2108.00908 [nucl-ex]].
%
%
\bibitem{ckp.nrcki.ru}
\url{http://ckp.nrcki.ru/}

\bibitem{govorun}
\url{http://hlit.jinr.ru/supercomputer_govorun/}
%
%
%
%
%
%
%
\end{thebibliography}
\end{document}